\documentclass{giq7}
\newcommand{\rd}{\mathrm{d}}

\theoremstyle{definition}

\usepackage{graphicx}
\makeindex

\begin{document}

\title[Variational Problems in Elastic Theory]
{VARIATIONAL PROBLEMS IN ELASTIC THEORY OF BIOMEMBRANES, SMECTIC-A
LIQUID CRYSTALS, AND CARBON RELATED STRUCTURES}

\author{Z. C. TU \MakeLowercase{and} Z. C. OU-YANG}
\address{Institute of Theoretical Physics, Academia Sinica,
P.O.Box 2735 Beijing 100080, China}

\begin{abstract}
After a brief introduction to several variational problems in the
study of shapes of thin structures, we deal with variational
problems on 2-dimensional surface in 3-dimensional Euclidian space
by using exterior differential forms and the moving frame method.
The morphological problems of lipid bilayers and stabilities of
cell membranes are also discussed. The key point is that the first
and the second order variations of the free energy determine
equilibrium shapes and mechanical stabilities of structures.
\end{abstract}

\maketitle

\section{Introduction}\label{sec-Intr}
The morphology of thin structures (always represented by a smooth
surface $M$ in this paper) is an old problem. First, we look back
on the history \cite{oytfm}. As early as in 1803, Plateau studied
a soap film attaching to a metallic ring when the ring passed
through soap water \cite{Plateau}. By taking the minimum of the
free energy $F=\lambda\int_M \rd A$, he obtained $H=0$, where
$\lambda$ and $H$ are the surface tension and mean curvature of
the soap film, respectively. From 1805 and 1806, Young
\cite{Young} and Laplace \cite{Laplace} studied soap bubbles. By
taking the minimum of the free energy $F=p\int \rd
V+\lambda\oint_M \rd A$, they obtained $H=p/2\lambda$, where $p$
is the osmotic pressure (pressure difference between outer and
inner sides) of a soap bubble and $V$ is the volume enclosed by
the bubble. We can only observe spherical bubbles because ``An
embedded surface with constant mean curvature in 3-dimensional
(3D) Euclidian space ($\mathbb{E}^3$) must be a spherical
surface'' \cite{Alexandrov}. In 1812, Poisson \cite{Poisson}
considered a solid shell and put forward the free energy
$F=\oint_M H^2 \rd A$. Its Euler-Lagrange equation is $\nabla^2
H+2H(H^2-K)=0$ \cite{Willmore}. Now the solutions to this equation
are called Willmore surfaces. In 1973, Helfrich recognized that
lipid bilayers could be regarded as smectic-A (SmA) liquid
crystals (LCs) at room temperature. Based on the elastic theory of
liquid crystals \cite{degennes}, he proposed the curvature energy
per unit area of the bilayer
\begin{align}\label{Helfrich} \mathcal{E}_{lb}=(k_c/2)(2H+c_0)^2+\bar{k}K, \end{align}
where $k_c$ and $\bar{k}$ are elastic constants. $K$ and $c_0$ are
Gaussian curvature and spontaneous curvature of the lipid bilayer,
respectively. Starting with Helfrich's curvature energy
(\ref{Helfrich}), the morphology of lipid vesicles has been deeply
understood \cite{Lipowsky,oybook,Seifert}. Especially, the free
energy is expressed as $F=p \int \rd V+\oint_M
(\lambda+\mathcal{E}_{lb}) \rd A$ for lipid vesicles, and the
corresponding Euler-Lagrange equation is \cite{oyprl}:
\begin{align}p-2\lambda H+k_c\nabla^2(2H)
+k_c(2H+c_0)(2H^2-c_0H-2K)=0.\label{shapelbc}
\end{align}
For an open lipid bilayer with a free edge $C$, the free energy is
expressed as $F=\int_M (\lambda+\mathcal{E}_{lb}) \rd
A+\gamma\oint_C \rd s$, where $\gamma$ is the line tension of the
edge. The corresponding Euler-Lagrange equations are as follows
\cite{Capovilla,tzcpre}:
\begin{align}
k_{c}(2H+c_{0})(2H^{2}-c_{0}H-2K)-2\lambda H+k_{c}\nabla ^{2}(2H)
&=0\label{openby1}\\
\left. \left[ k_{c}(2H+c_{0})+\bar{k}k_n\right]\right\vert _{C} &=0\label{openby2} \\
\left. \left[ -2k_{c}\frac{\partial
H}{\partial\mathbf{e}_2}+\gamma k_n+\bar{k}
\frac{d\tau_g}{ds}\right]\right\vert _{C} &=0\\
\left. \left[ \frac{k_{c}}{2}(2H+c_{0})^{2}+\bar{k}K+\lambda
+\gamma k_{g}\right]\right\vert _{C}&=0,\label{openby4}
\end{align}
where $k_n$, $k_g$, and $\tau_g$ are normal curvature, geodesic
curvature, and geodesic torsion of the boundary curve. The unit
vector $\mathbf{e}_2$ (see also Fig.~\ref{fig1}) is perpendicular
to tangent vector of edge $C$ and normal vector of surface $M$.
Above four equations are called the shape equation and boundary
conditions of open lipid bilayers. The boundary conditions are
available for open lipid bilayers with more than one edge because
the edge in our derivation \cite{tzcpre} is a general one.

Secondly, we turn to the puzzle about the formation of focal conic
structures in SmA LCs. As we imagine, the configuration of minimum
energy in SmA LCs is a flat layer structure. But Dupin cyclides
are usually formed when LCs cool from isotropic phase to SmA phase
in the experiment \cite{Friedel}. Why the cyclides are preferred
to other geometrical structures under the preservation of the
interlayer spacing \cite{Bragg}? This phenomenon can be understood
by the concept that the Gibbs free energy difference between
isotropic and SmA phases must be balanced by the curvature elastic
energy of SmA layers \cite{Naitoprl}. The total free energy
includes curvature energy, volume energy and surface energy. It is
expressed formally as $F=\oint \mathcal{E}(H,K,t)\rd A$, where $t$
is the thickness of the focal conic domain; $H$ and $K$ are mean
curvature and Gaussian curvature of the inmost layer surface,
respectively. The Euler-Lagrange equations corresponding to the
free energy are as follows \cite{Naitopre}:
\begin{align}
\oint (\partial\mathcal{E}/\partial t)\rd A & =0\\
(\nabla^2/2+2H^2-K)\partial \mathcal{E}/\partial
H+(\nabla\cdot\tilde{\nabla}+2KH)\partial \mathcal{E}/\partial
K-2H\mathcal{E}& =0.
\end{align}
Solving both equations can give good explanation to focal conic
domains \cite{Naitopre}. The new operator $\tilde{\nabla}$ can be
found in in the appendix of Ref.~\cite{tzcjpa}.

Thirdly, let us see carbon related structures. There are three
typical structures composed of carbon atoms: Buckyball (C$_{60}$),
single-walled carbon nanotube (SWNT), and carbon torus. In the
continuum limit, we derive the curvature energy \cite{oysu,tzcprb}
of single graphitic layer $E=\int \left[\frac{1}{2} k_{c}
(2H)^2+\bar{k} K \right] \rd A$ from the lattice model
\cite{Lenosky}, where $k_{c}$ and $\bar{k}$ are elastic constants.
The total free energy of a graphite layer is $F=\int
\left[\frac{1}{2} k_{c} (2H)^2+\bar{k} K \right] \rd A
+\lambda\int \rd A$, where $\lambda$ is the surface energy per
unit area for graphite. Please note that the surface energy per
unit area for solid structures is not as a constant quantity as
the surface tension for fluid membranes. The Euler-Lagrange
equation corresponding to the free energy is $\nabla^2
H+2H(H^2-K)-\lambda H/k_c=0$. C$_{60}$ and carbon torus can be
understood with $\lambda=0$, while SWNT satisfies
$R^2=k_c/2\lambda$, where $R$ is its radius.

The rest of this paper is organized as follows. In
Sec.~\ref{sec-var2d}, we show how to derive the Euler-Lagrange
equation from the free energy functional by using exterior
differential forms. The method is developed in Refs.~\cite{tzcpre}
and \cite{tzcjpa}, which might be equivalent to the work by
Griffiths \cite{Griffiths} in essence. But it is more convenient
to apply in variational problems on 2D surface in $\mathbb{E}^3$.
In Sec.~\ref{sec-lb}, we give several analytic solutions to the
shape equation of lipid vesicles, and to the shape equation and
boundary conditions of open lipid bilayers. In Sec.~\ref{sec-cm},
we discuss the elasticity and stability of cell membranes. A brief
summary is given in the last section.

\section{Variational problems on 2D surface}\label{sec-var2d}
Many variational problems are shown in Introduction. Here we deal
with them by using exterior differential forms.

\begin{figure}[htp!]
\includegraphics[width=5cm]{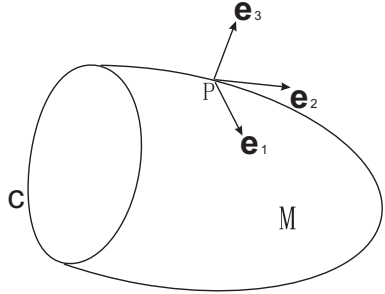}
\caption{\label{fig1}A surface $M$ with an edge $C$.
\{$\mathbf{e}_1,\mathbf{e}_2,\mathbf{e}_3$\} is an orthonormal
frame.}
\end{figure}

Let us consider a surface $M$ with an edge $C$ as shown in
Fig.~\ref{fig1}. At every point P in the surface, we can choose a
right-handed, orthonormal frame
\{$\mathbf{e}_1,\mathbf{e}_2,\mathbf{e}_3$\} with $\mathbf{e}_3$
being the normal vector. For a point in curve $C$, $\mathbf{e}_1$
is the tangent vector of $C$ such that $\mathbf{e}_2$ points to
the side that the surface is located in.

The differential of the frame is expressed as
\begin{align}
\rd\mathbf{r}&=\omega_1\mathbf{e}_1+\omega_2\mathbf{e}_2\\
\rd\mathbf{e}_i&=\omega_{ij}\mathbf{e}_j,\quad (i=1,2,3)
\end{align}
where $\omega_1$, $\omega_2$, $\omega_{ij}=-\omega_{ji}$
($i,j=1,2,3$) are 1-forms. Please note that the repeated subindex
represents the summation from 1 to 3, unless otherwise specified
in this paper.

The structure equations of the surface are as follows:
\begin{align}
\rd\omega_1&=\omega_{12}\wedge\omega_2\\
\rd\omega_2&=\omega_{21}\wedge\omega_1\\
\omega_{13}&=a\omega_{1}+b\omega_{2},\omega_{23}=b\omega_{1}+c\omega_{2}\\
\rd\omega_{ij}&=\omega_{ik}\wedge\omega_{kj}\quad (i,j=1,2,3),
\end{align}
where $a$, $b$, $c$ are related to the curvatures with $2H=a+c$
and $K=ac-b^2$.

The variation of the frame is denoted by
\begin{align}
\delta
\mathbf{r}&=\delta_1\mathbf{r}+\delta_2\mathbf{r}+\delta_3\mathbf{r}\\
\delta_i\mathbf{r}&=\Omega_{i}\mathbf{e}_{i}\quad (i=1,2,3)\label{deltair}\\
\delta_l \mathbf{e}_{i}&=\Omega _{lij}\mathbf{e}_{j}, (i,l=1,2,3),
\end{align}
with $\Omega _{lij}=-\Omega _{lji}$ ($i,j,l=1,2,3$). In
equation~\eqref{deltair}, the repeated subindex does not represent
summation. It is easy to prove that the operator $\delta_l
(l=1,2,3)$ has the similar properties with the partial
differential while the operator $\delta$ has the similar
properties with the total differential operator \cite{tzcjpa}.

Using $\rd\delta_l \mathbf{r}=\delta_l \rd\mathbf{r}$ and
$\rd\delta_l \mathbf{e}_i=\delta_l \rd\mathbf{e}_i$, we obtain
variational equations of the frame as follows \cite{tzcjpa}:
\begin{align}
\delta_1 \omega _{1} &=\rd\Omega _{1}-\omega _{2}\Omega _{121}\\
\delta_1 \omega _{2} &=\Omega _{1}\omega _{12}-\omega _{1}\Omega _{112}\\
\Omega _{113}&=a\Omega _{1},\quad\Omega _{123}=b\Omega _{1};
\end{align}
\begin{align}
\delta _{2}\omega _{1} &=\Omega _{2}\omega _{21}-\omega _{2}\Omega _{221} \\
\delta _{2}\omega _{2} &=\rd\Omega _{2}-\omega _{1}\Omega _{212}\\
\Omega _{213} &=b\Omega _{2},\quad\Omega _{223}=c\Omega _{2};
\end{align}
\begin{align}
\delta_3 \omega _{1} =\Omega _{3}\omega _{31}-\omega _{2}\Omega
_{321}\\
\delta_3 \omega _{2} =\Omega _{3}\omega _{32}-\omega _{1}\Omega
_{312}\\
\rd\Omega _{3} =\Omega _{313}\omega _{1}+\Omega _{323}\omega _{2};
\end{align}
\begin{align}\delta_l \omega _{ij}=\rd\Omega _{lij}+\Omega _{lik}\omega
_{kj}-\omega _{ik}\Omega _{lkj}.
\end{align}

Using them, we can prove that
\begin{align}
\delta_1(\mathcal{E}\omega _{1}\wedge\omega _{2}) &=\rd (\mathcal{E}\omega _{2}\Omega _{1})\label{lemma1}\\
\delta_2(\mathcal{E}\omega _{1}\wedge\omega _{2})
&=-\rd(\mathcal{E}\omega _{1}\Omega
_{2})\\
\delta_3 (\omega _{1}\wedge\omega _{2})&=-2H\Omega _{3}\omega
_{1}\wedge\omega _{2}\\
\delta_3 (2H) \omega _{1}\wedge\omega _{2}&=2(2H^{2}-K)\Omega
_{3}\omega _{1}\wedge\omega _{2}+\rd\ast \rd\Omega_3\\
\delta_3 K \omega _{1}\wedge\omega _{2}&=2KH\Omega_3\omega
_{1}\wedge\omega
_{2}+\rd\tilde{\ast}\tilde{\rd}\Omega_{3},\label{lemma2}
\end{align}
where $\mathcal{E}$ is a function of $2H$ and $K$. $\ast$ is Hodge
star operator \cite{Westenholz} satisfying the following
properties: (i) $\ast f=f\omega_1\wedge\omega_2$ for scalar
function $f$; (ii) $\ast\omega_1=\omega_2$,
$\ast\omega_2=-\omega_1$. $\tilde{\ast}$ and $\tilde{\rd}$ are new
operators defined in Ref.~\cite{tzcjpa} that satisfy: (i) If
$df=f_1\omega_1+f_2\omega_2$, then $\tilde{\rd}
f=f_1\omega_{13}+f_2\omega_{23}$; (ii)
$\tilde{\ast}\omega_{13}=\omega_{23},\tilde{\ast}\omega_{23}=-\omega_{13}$;
(iii)
$\int_M(f\rd{\ast}\tilde{\rd}h-h\rd{\ast}\tilde{\rd}f)=\int_{\partial
M}(f{\ast} \tilde{\rd}h-h{\ast} \tilde{\rd}f)$,
$\int_M(f\rd\tilde{\ast}\tilde{\rd}h-h\rd\tilde{\ast}\tilde{\rd}f)=\int_{\partial
M}(f\tilde{\ast} \tilde{\rd}h-h\tilde{\ast} \tilde{\rd}f)$ for any
smooth functions $f$ and $h$ on $M$. \footnote{These two
expressions are very similar to the second Green identity. Their
proof can be found in Lemma 2.1 of Ref.~\cite{tzcjpa}.}

Now, we consider the variational problem on closed surface. In
this case, the general functional is expressed as:
\begin{align}\label{functves}
F=\oint_M \mathcal{E}(2H,K)\rd A+p\int_V \rd V.
\end{align}
Using Stokes theorem and the variational equations of the frame,
we can prove that
\begin{align}
\delta_1\int_V \rd V&=\delta_2\int_V \rd V=0\\
\delta_3\int_V \rd V&=\oint_M \Omega_{3}\rd A.
\end{align}
Combining them with equations~\eqref{lemma1}--\eqref{lemma2}, we
have $\delta_1 F=\delta_2 F=0$, and the Euler-Lagrange equation
corresponding to functional \eqref{functves}:
\begin{align}
\left[\left(\nabla^2+4H^2-2K\right)\frac{\partial}{\partial
(2H)}+\left(\nabla\cdot\tilde{\nabla}+2KH\right)\frac{\partial}{\partial
K}-2H\right]\mathcal{E}+ p=0.\label{eulclos}
\end{align}

Next, we consider the variational problem on open surface with an
edge $C$. In this case, the general functional is expressed as:
\begin{align}\label{functolb}
F=\int_M \mathcal{E}(2H,K)\rd A+\oint_C\Gamma(k_n,k_g)\rd s.
\end{align}
Similarly, we derive its Euler-Lagrange equations as
\begin{align}
&(\nabla ^{2}+4H^{2}-2K)\frac{\partial \mathcal{E}}{\partial
(2H)}+(\nabla
\cdot \tilde{\nabla}+2KH)\frac{\partial \mathcal{E}}{\partial K}-2H\mathcal{E%
}=0\label{euleropen1}\\
&\mathbf{e}_{2}\cdot \nabla \left[ \frac{\partial \mathcal{E}}{\partial (2H)%
}\right] +\mathbf{e}_{2}\cdot \tilde{\nabla}\left( \frac{\partial \mathcal{E}%
}{\partial K}\right) -\frac{d}{ds}\left(\tau_g \frac{\partial \mathcal{E}}{%
\partial K}\right) +\frac{d^{2}}{ds^{2}}\left( \frac{\partial \Gamma }{%
\partial k_{n}}\right) +\frac{\partial \Gamma }{\partial k_{n}}%
(k_{n}^{2}-\tau _{g}^{2})\nonumber \\
&\quad +\tau _{g}\frac{d}{ds}\left( \frac{\partial \Gamma }{\partial k_{g}}%
\right) +\frac{d}{ds}\left( \tau _{g}\frac{\partial \Gamma }{\partial k_{g}}%
\right) -\left. \left( \Gamma -\frac{\partial \Gamma }{\partial k_{g}}%
k_{g}\right) k_{n}\right\vert _{C}=0 \\
&-\frac{\partial \mathcal{E}}{\partial (2H)}-k_{n}\frac{\partial \mathcal{E}%
}{\partial K}+\frac{\partial \Gamma }{\partial k_{g}}k_{n}-\left. \frac{%
\partial \Gamma }{\partial k_{n}}k_{g}\right\vert _{C}=0 \\
&\frac{d^{2}}{ds^{2}}\left( \frac{\partial \Gamma }{\partial
k_{g}}\right) +K\frac{\partial \Gamma }{\partial
k_{g}}-k_{g}\left( \Gamma -\frac{\partial \Gamma }{\partial
k_{g}}k_{g}\right) +2(k_{n}-H)k_{g}\frac{\partial \Gamma
}{\partial k_{n}}\nonumber \\
&\quad -\tau_g\frac{d}{ds}\left( \frac{\partial \Gamma }{\partial
k_{n}}\right)
\left. -\frac{d}{ds}\left( \tau_g\frac{\partial \Gamma }{\partial k_{n}}\right) -%
\mathcal{E}\right\vert _{C}=0.\label{euleropen2}
\end{align}

In special cases, above equations \eqref{eulclos},
\eqref{euleropen1}--\eqref{euleropen2} are degenerated into
different equations mentioned in Introduction. For example, if
taking $\mathcal{E}=\mathcal{E}_{lb}+\lambda$ and $\Gamma=\gamma$,
These equations are degenerated into equations
\eqref{openby1}--\eqref{openby4}, respectively.

\section{Morphology of lipid bilayers}\label{sec-lb}
There are three typical solutions to shape equation
\eqref{shapelbc}: Sphere, torus, and biconcave vesicle.

\begin{figure}[htp!]
\includegraphics[width=5cm]{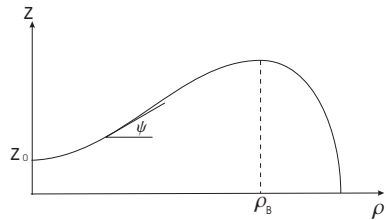}
\caption{\label{fig2}Schematic figure of the contour of biconcave
vesicle in the first quadrant.}
\end{figure}

First, a spherical vesicle with radius $R$ satisfies equation
\eqref{shapelbc} if $p R^2+2\lambda R-k_cc_0(2-c_0R)=0$ is valid.
Secondly, Ou-Yang used equation \eqref{shapelbc} to predict that a
torus with the radii of two generating circles $r$ and $R$
satisfying $R/r=\sqrt{2}$ should be observed in lipid systems
\cite{oypra}. This striking prediction has been confirmed
experimentally by three groups \cite{Lin94,Mutz91,Rudolph}.
Thirdly, the first exact axisymmetric solution with biconcave
shape, as shown in Fig.~\ref{fig2}, was found under the condition
of $p=\lambda=0$ as \cite{Naitopre93}:
\begin{align}z&=z_0+\int_0^\rho \tan\psi(\rho')d\rho'\\
\sin\psi(\rho)&=-c_0\rho\ln(\rho/\rho_B), c_0>0,
\end{align}
where $z(\rho)$ is the contour of the cross-section. $z$ axis is
the rotational axis, and $\psi(\rho)$ the tangent angle of the
contour at distance $\rho$. This solution can explain the classic
physiological puzzle \cite{Fung}: Why the red blood cells of
humans are always in biconcave shape?

\begin{figure}[htp!]
\includegraphics[width=3cm]{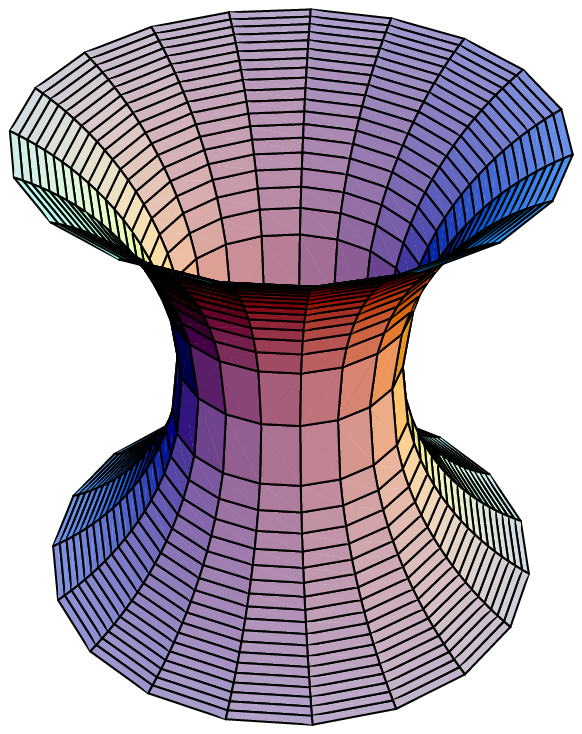}
\includegraphics[width=3.4cm]{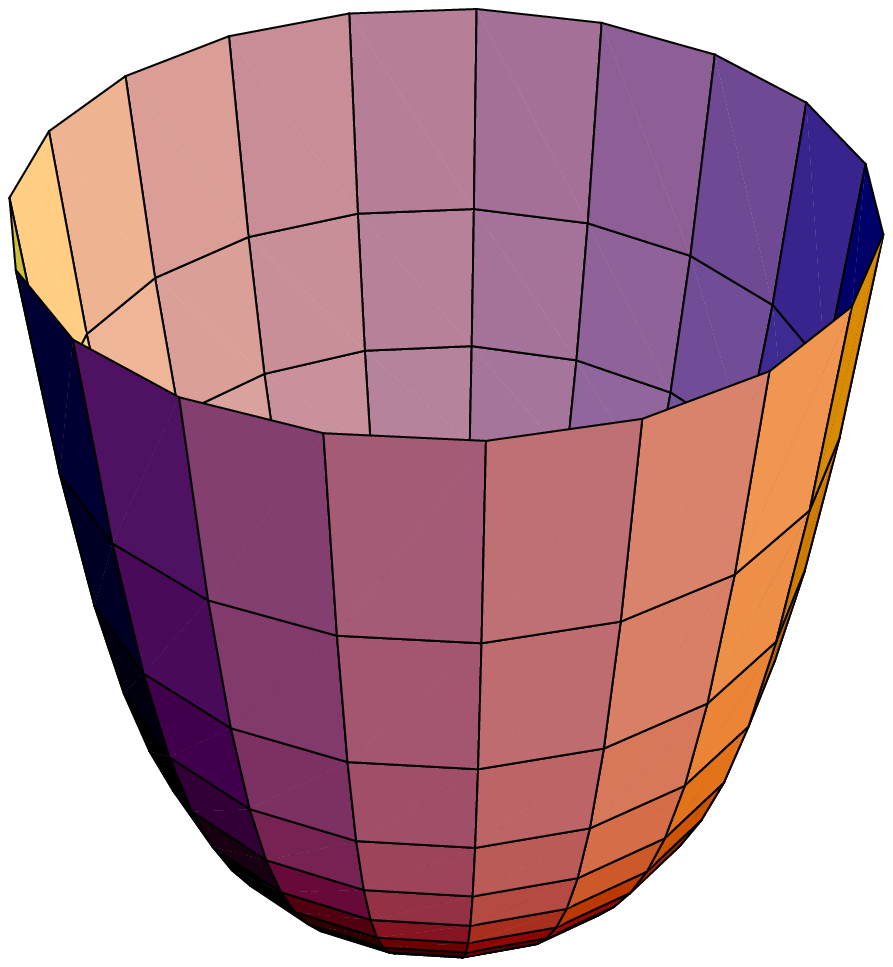}
\caption{\label{fig3}The central part of a torus (left) and a
cup-like membrane (right).}
\end{figure}

To the shape equation \eqref{openby1} and boundary conditions
\eqref{openby2}--\eqref{openby4} of open lipid bilayers, we can
find two analytical solutions \cite{tzcpre}: One is the central
part of a torus and another is a cup-like membrane shown in
Fig.~\ref{fig3}. Numerical method and solutions to these equations
can be find in Ref.~\cite{Umeda}.

\section{Elasticity and stability of cell membranes}\label{sec-cm}
A cell membrane is simplified as lipid bilayer plus membrane
skeleton. The skeleton is a cross-linking protein network and
joint to the bilayer at some points. We know that the
cross-linking polymer structure also exists in rubber at molecular
levels. Thus we can transplant the theory of rubber elasticity
\cite{Treloar} to describe the membrane skeleton.

Based on Helfrich's theory and physics of rubber elasticity, the
free energy of a closed cell membrane can be expressed as
\cite{tzcjpa}:
\begin{align}
F=\oint_M(\mathcal{E}_d+\mathcal{E}_H)\rd A+p\int_V\rd V,
\end{align}
with $\mathcal{E}_H=(k_c/2)(2H+c_0)^2+\lambda$ and
$\mathcal{E}_d=(k_d/2)[(2J)^{2}-Q]$, where $2J=\varepsilon
_{11}+\varepsilon _{22}$, $Q=\varepsilon _{11}\varepsilon
_{22}-\varepsilon _{12}^2$. Here $\varepsilon _{ij}$ ($i,j=1,2$)
represents the in-plane strain of the membrane. $k_d$ is the
elastic constant representing the entropic elasticity
\cite{gennes} of membrane skeleton.

Using the method in Sec.~\ref{sec-var2d}, we obtain the shape
equation and the in-plane strain equations of the cell membrane as
\cite{tzcjpa}:
\begin{align}
&p-2H(\lambda+k_{d}J)+k_{c}(2H+c_{0})(2H^{2}-c_{0}H-2K)\nonumber\\&\qquad+k_{c}\nabla
^{2}(2H)-\frac{k_{d}}{2}(a\varepsilon _{11}+2b\varepsilon
_{12}+c\varepsilon _{22}) =0\\
&k_d[-\rd(2J)\wedge \omega _{2}-\frac{1}{2}(\varepsilon
_{11}\rd\omega _{2}-\varepsilon _{12}\rd\omega
_{1})+\frac{1}{2}\rd(\varepsilon _{12}\omega _{1}+\varepsilon
_{22}\omega _{2})] =0
\\
&k_d[\rd(2J)\wedge \omega _{1}-\frac{1}{2}(\varepsilon
_{12}\rd\omega _{2}-\varepsilon _{22}\rd\omega
_{1})-\frac{1}{2}\rd(\varepsilon _{11}\omega _{1}+\varepsilon
_{12}\omega _{2})] =0.
\end{align}
An obvious solution is the spherical cell membrane with homogenous
strains: $\varepsilon_{11}=\varepsilon_{22}=\varepsilon$ (a
constant) and $\varepsilon_{12}=0$. The radius $R$ of the sphere
must satisfy
\begin{align}\label{spherecm}
pR^{2}+(2\lambda +3k_{d}\varepsilon )R+k_{c}c_{0}(c_{0}R-2) =0.
\end{align}

Now we will show the biological function of membrane skeleton by
discussing the mechanical stability of a spherical cell membrane.
Using Hodge decomposed theorem \cite{Westenholz}, $\Omega_1$ and
$\Omega_2$ can be expressed as $\Omega _{1}\omega _{1}+\Omega
_{2}\omega _{2}=d\Omega +\ast d\chi$ by two scalar functions
$\Omega$ and $\chi$. Through complex calculations, we obtain the
second order variation of the free energy for spherical membrane
$\delta ^{2}\mathcal{F}=G_{1}+G_{2}$, where
\begin{align}
G_{1} &=\oint_{M}\Omega
_{3}^{2}\{3k_{d}/R^{2}+(2k_{c}c_{0}/R^{3})+p/R\}\rd A\nonumber\\
&+\oint_{M}\Omega _{3}\nabla
^{2}\Omega _{3}\{k_{c}c_{0}/R+2k_{c}/R^{2}+pR/2\}\rd A \nonumber\\
&+\oint_{M}k_{c}(\nabla ^{2}\Omega _{3})^{2}\rd A+\frac{3k_{d}}{R}%
\oint_{M}\Omega _{3}\nabla ^{2}\Omega \rd A\nonumber\\
&+k_{d}\oint_{M}\left( \nabla ^{2}\Omega \right) ^{2}\rd
A+\frac{k_{d}}{2R^{2}}\oint_{M}\Omega \nabla ^{2}\Omega \rd A,
\end{align}
and $G_{2}=(k_{d}/4)\oint_{M}\nabla ^{2}\chi (\nabla ^{2}\chi
+2\chi /R^{2})\rd A$. Because $G_2$ is positive definite, we
merely need to discuss $G_1$. $\Omega_{3}$ and $\Omega$ in the
expression of $G_1$ are arbitrary functions defined in a sphere
and can be expanded by spherical harmonic functions \cite{Wangzx}:
$\Omega _{3} =\sum_{l=0}^{\infty
}\sum_{m=-l}^{m=l}a_{lm}Y_{lm}(\theta,\phi)$ and $\Omega
=\sum_{l=0}^{\infty }\sum_{m=-l}^{m=l}b_{lm}Y_{lm}(\theta,\phi)$
with $a_{lm}^{*} =(-1)^{m}a_{l,-m}$ and $b_{lm}^{*}
=(-1)^{m}b_{l,-m}$. Considering \eqref{spherecm}, we write $G_1$
in a quadratic form:
\begin{align}
G_{1} &=\sum_{l=0}^{\infty
}\sum_{m=0}^{l}2|a_{lm}|^{2}%
\{3k_{d}+[l(l+1)-2][l(l+1){k}_{c}/R^{2}-{k}_{c}{c}_{0}/R- pR/2]\}\nonumber\\
&-\sum_{l=0}^{\infty
}\sum_{m=0}^{l}\frac{3k_{d}}{R}l(l+1)(a_{lm}^{\ast
}b_{lm}+a_{lm}b_{lm}^{\ast })\nonumber\\ &+\sum_{l=0}^{\infty }\sum_{m=0}^{l}\frac{k_{d}}{%
R^{2}}\left[ 2l^{2}(l+1)^{2}-l(l+1)\right] |b_{lm}|^{2}.
\end{align}

It is easy to prove that, if $p<p_{l}=\frac{3k_{d}}{\left[ 2l(l+1)-1\right] R}+\frac{2{k}_{c}[l(l+1)-{c}_{0}R]%
}{R^{3}}\quad(l=2,3,\cdots)$, then $G_1$ is positive definite.
Thus we must take the minimum of $p_l$ to obtain the critical
pressure \footnote{In Ref.~\cite{tzcjpa}, we ignore the effect of
in-plane modes $\Omega_1$ and $\Omega_2$ on the critical pressure
and obtain the invalid value.}:
\begin{align}p_{c}=\min \{p_{l}\}=\left\{
\begin{array}{c}
\frac{3k_{d}}{11R}+\frac{2k_c[6-c_0R]}{R^{3}}<\frac{k_c[23-2c_0R]}{%
R^{3}},\quad (3k_{d}R^{2}<121k_c) \\
\frac{2\sqrt{3k_{d}k_c}}{R^{2}}+\frac{k_c}{R^{3}}(1-2c_0R),\quad
(3k_{d}R^{2}>121k_c).
\end{array}%
\right.\label{criticalpcm}
\end{align}
Taking typical data of cell membrane, $k_c\sim 20k_BT$
\cite{Duwe}, $k_d\sim 6\times 10^{-4}k_BT/nm^2$ \cite{Lenormand},
$h\sim 4nm$, $R\sim 1\mu m$, $c_0R\sim 1$, we have $p_c\sim 2$ Pa
from equation~\eqref{criticalpcm}. If not considering membrane
skeleton, that is $k_d=0$, we obtain $p_c\sim 0.2$ Pa. Therefore,
membrane skeleton enhances the mechanical stability of cell
membranes, at least for spherical shape.

\section{Summary}\label{sec-sum}
In above discussion, we introduce several problems in the
elasticity of biomembranes, smectic-A liquid crystal, and carbon
related structures. We deal with these variational problems on 2D
surface by using exterior differential forms. Elasticity and
stability of lipid bilayers and cell membranes are calculated and
compared with each other. It is shown that membrane skeleton
enhances the mechanical stability of cell membranes.

\section*{Acknowledgement}
ZCT would like to thank the useful discussion and kind help of
Prof. Y. S. Cho, T. Ivey, I. Mladenov, V. M. Vassilev, O.
Yampolsky, and I. Zlatanov during this conference.


\begin{thebibliography}{99}
\bibitem{Alexandrov}Alexandrov A. D., \emph{Uniqueness Theorems for Surfaces in the Large}, Amer. Math. Soc. Transl.,
\textbf{21} (1962) 341--416.

\bibitem{Bragg}Bragg W., \emph{Liquid Crystals} Nature \textbf{133} (1934)
445.

\bibitem{Capovilla}Capovilla R., Guven J. and Santiago J.A., \emph{Lipid Membranes with an Edge}, Phys. Rev. E \textbf{66}
(2002) 21607.

\bibitem{degennes}de Gennes P. G., \emph{The Physics of Liquid Crystals}, Clarendon Press, Oxford,
1975.

\bibitem{gennes}de Gennes P. G., \emph{Scaling Concepts in Polymer Physics}, Cornell University Press, New York, 1979.

\bibitem{Duwe}Duwe H P, Kaes J and Sackmann E, \emph{Bending Elastic-Moduli of Lipid Bilayers--Modulation by Solutes}, \textit{J. Phys. Fr.} \textbf{51} (1990) 945--962.

\bibitem{Friedel}Friedel G. and Grandjean F., \emph{Observations Geometriques sur les Liquides a Conique Focales}, Bull. Soc. fr.
Miner. \textbf{33} (1910) 409--465.

\bibitem{Fung}Fung Y. C. and Tong P., \emph{Theory of the Sphering of Red Blood Cells} Biophys. J. \textbf{8} (1968) 175--198.

\bibitem{Griffiths}Griffiths P., \emph{Exterior Differential Systems
and the Calculus of Variations}, Birkh\"{a}user, Boston, 1983.

\bibitem{Helfrich}Helfrich W., \emph{Elastic Properties of Lipid Bilayers--Theory and Possible Experiments}, Z. Naturforsch. C
\textbf{28} (1973) 693.

\bibitem{Laplace}Laplace P. S., \emph{Trait\'{e} de M\'{e}canique C\'{e}leste}, Gauthier-Villars,
Paris, 1839.

\bibitem{Lenormand}Lenormand G., H\'{e}non S., Richert A., Sim\'{e}on J., and
Gallet F., \emph{Direct Measurement of the Area Expansion and
Shear Moduli of the Human Red Blood Cell Membrane Skeleton},
Biophys. J. \textbf{81} (2001) 43--56.

\bibitem{Lenosky}Lenosky T., Gonze X., Teter M. and Elser V., \emph{Energetics of Negatively Curved Graphitic Carbon}. Nature \textbf{355} (1992) 333--335.

\bibitem{Lin94}Lin Z., Hill R. M., Davis H. T., Scriven L. E. and Talmon Y., \emph{Cryo Transmission Electron Microscopy Study
of Vesicles and Micelles in Siloxane Surfactant Aqueous
Solutions}, Langmuir \textbf{10} (1994) 1008--1011.

\bibitem{Lipowsky}Lipowsky R., \emph{The Conformation of Membranes}, Nature \textbf{349} (1991) 475--481.

\bibitem{Mutz91}Mutz M. and  Bensimon D., \emph{Observation of Toroidal Vesicles}, Phys. Rev. A
\textbf{43} (1991) 4525--4527.

\bibitem{Naitoprl}Naito H., Okuda M. and Ou-Yang Z. C., \emph{Equilibrium Shapes Of Smectic-A Phase Grown from Isotropic
Phase}, Phys. Rev. Lett. \textbf{70} (1993) 2912--2915.

\bibitem{Naitopre93}Naito H., Okuda M. and Ou-Yang Z. C., \emph{Counterexample to Some Shape Equations for Axisymmetric
Vesicles}, Phys. Rev. E (1993) 2304--2307.

\bibitem{Naitopre}Naito H., Okuda M. and Ou-Yang Z. C., \emph{Preferred Equilibrium Structures of a Smectic-A Phase Grown from an Isotropic Phase: Origin of Focal Conic Domains}, Phys. Rev. E \textbf{52} (1995) 2095-2098.

\bibitem{oypra}Ou-Yang Z. C., \emph{Anchor Ring-Vesicle Membranes}, Phys. Rev. A \textbf{41} (1990)
4517--4520.

\bibitem{oytfm}Ou-Yang Z. C., \emph{Elastic Theory of Biomembranes}, Thin Solid Films \textbf{393}
(2001) 19--23.

\bibitem{oyprl}Ou-Yang Z. C. and Helfrich W., \emph{Instability and Deformation of a Spherical Vesicle by
Pressure}, Phys. Rev. Lett. \textbf{59} (1987) 2486--2488.

\bibitem{oysu}Ou-Yang Z. C., Su Z. B. and Wang C. L., \emph{Coil Formation in Multishell Carbon Nanotubes: Competition between Curvature Elasticity and Interlayer
Adhesion}, Phys. Rev. Lett. \textbf{78} (1997) 4055--4058.

\bibitem{oybook}Ou-Yang Z. C., Liu J. X. and Xie Y. Z., \emph{Geometric Methods in the Elastic Theory of Membranes
in Liquid Crystal Phases}, World Scientific, Singapore, 1999.

\bibitem{Plateau}Plateau J., \emph{Statique Exp\'{e}rimentale et Th\'{e}orique des Liquides Soumis aux Seules Forces Mol\'{e}culaires}, Gauthier-Villars, Paris, 1873.

\bibitem{Poisson}Poisson S. D., \emph{Trait\'{e} de M\'{e}canique}, Bachelier, Paris, 1833.

\bibitem{Rudolph}Rudolph A. S., Ratna B. R. and Kahn B., \emph{Self-Assembling Phospholipid Filaments}, Nature,
\textbf{352} (1991) 52--55.

\bibitem{Seifert}Seifert U., \emph{Configurations of Fluid Membranes and Vesicles}, Adv. Phys. \textbf{46} (1997) 13--137.

\bibitem{Treloar}Treloar L. R. G., \emph{The Physics of Rubber Elasticity}, Clarendon Press, Oxford,
1975.

\bibitem{tzcprb}Tu Z. C. and Ou-Yang Z. C., \emph{Single-Walled and Multiwalled Carbon Nanotubes Viewed as Elastic Tubes With the Effective Young'S Moduli Dependent on Layer
Number}, Phys. Rev. B \textbf{65} (2002) 233407.

\bibitem{tzcpre}Tu Z. C. and Ou-Yang Z. C., \emph{Lipid Membranes with Free Edges}, Phys. Rev. E \textbf{68} (2003) 61915.

\bibitem{tzcjpa}Tu Z. C. and Ou-Yang Z. C., \emph{A Geometric Theory on the Elasticity of Bio-membranes}, J. Phys. A: Math. Gen. \textbf{37} (2004) 11407--11429.

\bibitem{Umeda}Umeda T., Suezaki Y., Takiguchi K. and Hotani H., \emph{Theoretical Analysis of Opening-up Vesicles with Single and Two Holes}, Phys. Rev. E \textbf{71} (2005) 11913.

\bibitem{Wangzx}Wang Z. X. and Guo D. R., \emph{Introduction to Special
Function}, Peking University Press, Beijing, 2000.

\bibitem{Westenholz}Westenholz C. V., \textit{Differential Forms in
Mathematical Physics}, North-Holland, Amsterdam, 1981.

\bibitem{Willmore}Willmore T. J., \emph{Total Curvature in Riemannian Geometry}, John Wiley \& Sons, New York, 1982.

\bibitem{Young}Young T., \emph{An Essay on the Cohesion of Fluids}, Philos. Trans. R. Soc. London, \textbf{95} (1805)
65--87.

\end{thebibliography}
\end{document}